\documentclass[paper]{JHEP3}
\usepackage {amsmath}
\usepackage{graphicx}
\usepackage{cancel}

\DeclareGraphicsExtensions{.eps}

\newcommand{\p}{\partial}

\newcommand{\half}{\tfrac{1}{2}}
\newcommand{\quater}{\tfrac{1}{4}}
\newcommand{\refeq}[1]{\stackrel{(\ref{#1})}{=}}
\newcommand{\refeqq}[2]{\stackrel{\scriptsize\begin{array}{c}(\ref{#1}) \\ (\ref{#2})\end{array}}{=}}
\newcommand{\nn}{\nonumber}

\newcommand{\EE}{{\cal E}}
\newcommand{\sectiono}[1]{\section{#1}\setcounter{equation}{0}}

\title{One entropy function to rule them all\ldots}

\author{Kevin Goldstein and Rudra P. Jena \\
  Tata Institute of Fundamental Research, Homi Bhabha Road, Mumbai 400 005,
  INDIA\\
  E-mail: \email{kevin,rpjena@theory.tifr.res.in}}

\preprint{{TIFR/TH/07-02} \\ {hep-th/0701221}}

\abstract{We study the entropy of extremal four dimensional black holes and five
  dimensional black holes and black rings is a unified framework using Sen's entropy
  function and dimensional reduction. The five dimensional black holes and black rings we
  consider project down to either static or stationary black holes in four dimensions. The
  analysis is done in the context of two derivative gravity coupled to abelian gauge
  fields and neutral scalar fields. We apply this formalism to various examples including
  $U(1)^3$ minimal supergravity. }


\begin{document}

\sectiono{Introduction}
\label{sec:intro}

The attractor mechanism has played a significant part in furthering our understanding of
black holes in string theory \cite{Ferrara:1995ih,Strominger:1996kf, Ferrara:1996dd}.  A
characteristic of extremal black holes, the mechanism fixes the near horizon metric and
field configuration of moduli independent of the moduli's asymptotic values.

While the original work was in the context of spherically symmetric supersymmetric
extremal black holes in $(3+1)$-dimensional ${\cal N}=2$ supergravity with two derivative
actions, the mechanism has been found to work in a much broader context. Examples of this
include non-supersymmetric theories, actions with higher derivative corrections, extremal
black holes in higher dimensions, rotating black holes and black rings
\cite{Goldstein:2005hq,Kallosh:2005ax,Tripathy:2005qp,Giryavets:2005nf,Goldstein:2005rr,Kallosh:2006bt,Kallosh:2006bx,Prester:2005qs,Alishahiha:2006ke,Sinha:2006yy,Chandrasekhar:2006kx,Maldacena:1997de,LopesCardoso:1998wt,LopesCardoso:1999cv,LopesCardoso:1999ur,LopesCardoso:1999xn,Mohaupt:2000mj,LopesCardoso:2000qm,LopesCardoso:2000fp,Dabholkar:2004yr,Dabholkar:2004dq,Sen:2004dp,Hubeny:2004ji,Bak:2005mt,Kraus:2005vz,Sen:2005wa,Sen:2005iz,Kraus:2005zm,Sahoo:2006rp,Kraus:2005gh,Sahoo:2006vz,Parvizi:2006uz,Alishahiha:2006jd,Ferrara:2006xx,Kallosh:1996vy,0605139,Astefanesei:2006dd,Kallosh:2006ib,Cardoso:2006cb,Morales:2006gm,Bellucci:2006ib,Sahoo:2006pm,Dabholkar:2006tb,Astefanesei:2006sy,Chandrasekhar:2006ic,Andrianopoli:2006ub,Mosaffa:2006gp,Cardoso:2006xz,D'Auria:2007ev,Cardoso:2007rg}.

In particular, by examining the BPS equations for black rings, \cite{Kraus:2005gh}, found
the attractor equations for supersymmetric extremal black rings. Motivated by the results
of \cite{Ferrara:1997tw,Goldstein:2005hq,Sen:2005wa}, which demonstrate the attractor
mechanism is independent of supersymmetry, we sought to show the attractor mechanism for
black rings with out recourse to supersymmetry by using the entropy function formalism of
\cite{Sen:2005wa}. We note that \cite{Dabholkar:2006za} have made use of the formalism for
studying small black rings.

Using the connection between four dimensional black holes and five dimensional black rings
in Taub-NUT \cite{Gaiotto:2005gf,Elvang:2005sa,Bena:2005ni} we construct the entropy
function for black rings. In fact we found that the same technique works for five
dimensional black holes.  This allows us to write down a single entropy function
describing both black holes and black rings --- one entropy function to rule them
all.\footnote{During the preparation of the paper, \cite{Cardoso:2007rg} appeared which
  carries out this analysis for a class of five dimensional rotating black holes.  }

In section~\ref{sec:thing} we discuss our set up and apply dimensional reduction from five
to four dimensions. In section~\ref{ads2s2u1} we study black holes and black rings whose
near horizon geometry have $AdS_2\times S^2\times U(1)$ symmetries. The $U(1)$ may be
non-trivially fibred. After dimensional reduction along the $U(1)$, we get an
$AdS_2\times S^2$ near horizon geometry. This class includes static black holes with
$AdS_2\times S^3$ horizons and black rings with $AdS_3\times S^2$ horizons. For the black
holes the $U(1)$ is fibred over the $S^2$ while for the ring we fibre over the $AdS_2$. We
specialise these examples to the case of Lagrangians with very special geometry and find
the BPS and non-BPS attractor equations.   In section~\ref{sec:gen} we
consider an $AdS_2\times U(1)^2$ horizon which projects down to an $AdS_2\times U(1)$. In
this case both $U(1)$'s may be non-trivially fibred.

\sectiono{Black thing entropy function and dimensional reduction}
\label{sec:thing}

We wish to apply the entropy function formalism \cite{Sen:2005wa, Sen:2005iz}, and its
generalisation to rotating black holes \cite{Astefanesei:2006dd}, to the five dimensional
black objects --- black rings and black holes. These objects are characterised by the
topology of their horizons. Black ring horizons have $S^2\times S^1$ topology while black
holes have $S^3$ topology.

We consider a five dimensional Lagrangian with gravity, $n_{v}$ Abelian gauge fields,
$\bar F^I$, $n_s$ neutral massless scalars, ${\bar X}^S$, and a Chern-Simons term:
\begin{equation}
  \label{acthd}
  S=\frac {1}{16\pi G_5}\int d^{5}x\sqrt{-\bar{g}}
  \bigg(
  {\bar R}
  -{\bar h }_{ST}(\vec {X})\p_\mu  {\bar X}^S\p^\mu {\bar X}^T
  -{\bar f}_{IJ}(\vec X ) \bar F^I_{\mu\nu}\bar F^{J\,\mu\nu}
  -{\bar c}_{IJK}{\bar \epsilon}^{\mu\nu\alpha\beta\gamma}\bar F^I_{\mu\nu}\bar F^J_{\alpha\beta}\bar A^K_\gamma\bigg),
\end{equation}
where ${\bar \epsilon}^{\mu\nu\alpha\beta\gamma}$ is the completely antisymmetric
{\em tensor} \/with ${\bar \epsilon}^{01234}=1/\sqrt{-\bar g}$. The gauge
couplings, ${\bar f}_{IJ}$, and the sigma model metric, ${\bar h}_{ST}$, are functions of
the scalars, ${\bar X}^S$, while the Chern-Simons coupling, ${\bar c}_{IJK}$, a completely
symmetric tensor, is taken to be independent of the scalars. The gauge field strengths are
related to the gauge potentials in the usual way: $\bar F^I=d\bar A^I$. We use bars to
distinguish $5D$ objects from the $4D$ ones which will appear after dimensional reduction.
We take the indices $\{I,\ldots,M\}$ to run over the $n_{v}$ $5D$ gauge fields and the
indices $\{S, T\}$ to run over the $n_{s}$ $5D$ scalars.

Since the Lagrangian density is not gauge invariant, we need to be slightly careful about
applying the entropy function formalism. Following \cite{Sahoo:2006rp} (who consider a
gravitational Chern-Simons term in three dimensions) we dimensionally reduce to a four
dimensional Lagrangian density which is gauge invariant. This allows us to find a reduced
Lagrangian and in turn the entropy function. As a bonus we will also obtain a relationship
between the entropy of four dimensional and five dimensional extremal solutions -- this is
the $4D$-$5D$ lift of \cite{Gaiotto:2005xt,Gaiotto:2005gf} in a more general context. The
relationship between the four and five dimensional charges is extensively discussed in
\cite{Gaiotto:2005xt,Gaiotto:2005gf}.

Assuming all the fields are independent of a compact direction $\psi$, we take the
ansatz\footnote{For simplicity, we will work in units in which the Taub-Nut modulus is set
  to $1$. Due to the attractor mechanism, the modulus will drop out of the final result. }
\begin{align}
  \label{sauron1b}
  ds^2&= w^{-1}g_{\mu\nu}dx^\mu dx^\nu
  + w^2(d\psi+A^0_\mu dx^\mu)^2,\\
  \bar A^I&= A^I_\mu dx^\mu+a^I(x^\mu)\left(d\psi+A^0_\mu dx^\mu\right),\label{sauron2b}\\
  {\bar X}^S &= {\bar X}^S(x^\mu).
  \label{sauron3b}
\end{align}
Whether space-time indices above run over $4$ or $5$ dimensions should be clear from the
context.  Performing dimensional reduction on $\psi$, the action becomes
\begin{equation}
  \label{eq:4d_action}
  S=\frac{1}{16\pi G_4}
  \int d^4 x \sqrt{-g}\Big( R -h_{st}(\vec\Phi)\p\Phi^s\p\Phi^t
  - f_{ij}(\vec \Phi)F^i_{\mu\nu}F^{j\;\mu\nu}
  - \tfrac{1}{2}\tilde{f}_{ij}(\vec{\Phi})\epsilon^{\mu\nu\alpha\beta}F^i_{\mu\nu}F^j_{\alpha\beta}\Big)
\end{equation}
where
\begin{equation}
  \label{eq:G4vsG5}
  \left(\int d\psi\right) G_4= G_5,
\end{equation}
$f _{ij}$ and $\tilde f _{ij}$ are
$(1+n_{v})\times (1+n_{v})$ matrices:
\begin{align}
  f_{ij}&=
  \bordermatrix{& 0 & J \cr 0 &\quater w^3+w{\bar f}_{LM}a^La^M & w {\bar f}_{JL}a^L \cr
    I & w {\bar f}_{IL}a^L & w {\bar f}_{IJ} \cr }
  \label{fij},\\
  \tilde{f}_{ij} &=
  \bordermatrix{& 0 & J \cr 0 & 2{\bar c}_{KLM}a^Ka^La^M & 3{\bar c}_{JKL}a^Ka^L \cr I &
    {3}{\bar c}_{IKL}a^Ka^L & 6{\bar c}_{IJK}a^K \cr  }
  \label{fijt},
  \intertext{and $h_{rs}$ is a diagonal $(1+n_{v}+n_s)\times(1+n_{v}+n_s)$ matrix:}
  h_{rs}&= \mathrm{diag}\left(
    \begin{array}{ccc}
      \frac{9}{2}w^{-2},  2w {\bar f}_{IJ}, {\bar h}_{RS}
    \end{array}
  \right)
  \label{hrs}.
\end{align}
The gauge indices, $\{i,j\}$, labeling the $(1+n_{v})$ four dimensional gauge fields, run
over $0,1,\ldots,n_{v}$. The additional gauge field, $A^0$ comes from the off-diagonal
part of the five dimensional metric while the remaining ones descend from the original
five dimensional gauge fields. The four dimensional gauge field strengths are given by
$F^i=(dA^0,dA^I)$ where the four dimensional gauge fields are given in terms of the $5D$
ones by (\ref{sauron2b}). The scalar indices, $\{r,s\}$, labelling the four dimensional
scalars, run over $(1+n_{v}+n_s)$ values. The first additional scalar, $w$, comes from the
size of the Kaluza-Klein circle. Then next set of $n_{v}$ scalars, which we label $a^I$,
come from the $\psi$-components of the five-dimensional gauge fields and become axions in
four dimensions. Lastly, the original $n_s$ five dimensional scalars, ${\bar X}^S$,
descend trivially.  We write the four dimensional scalars as,
$\Phi^{s}=(w,a^{I},X^{S})$. Finally, notice that the coupling,
$\tilde{f}_{ij}(\vec{\Phi})$, is built up out of the five-dimensional Chern-Simons
coupling and the axions. Details of the derivation of the form of $\tilde{f}_{ij}$ can
be found in Appendix~\ref{sec:dr}.

In the next two sections we shall consider what happens when the near-horizon geometries
have various symmetries.  Firstly, we will look at black holes and black rings with a
higher degree of symmetry, namely $AdS_2\times S_2\times U(1)$, where the $U(1)$ may be
non-trivially fibred. Upon dimensional reduction we obtain a static, spherically
symmetric, extremal black hole near-horizon geometry --- $AdS_2\times S_2$ --- for which
the analysis is much simpler. The entropy function formalism only involves algebraic
equations. After that we will look at black objects whose near horizon symmetries are
$AdS_2\times U(1)^2$ in five dimensions. Once again, the $U(1)$'s may be non-trivially
fibred. After dimensional reduction, we get an extremal, rotating, near horizon geometry
--- $AdS_2\times U(1)$ --- for which the entropy function analysis was performed in
\cite{Astefanesei:2006dd}. For this case, the formalism involves differential equations in
general.

\sectiono{Algebraic entropy function analysis}
\label{ads2s2u1}

In this section, we will construct and analyse the entropy function for five dimensional
black holes and black rings sitting in Taub-NUT space with $AdS_2\times S_2\times U(1)$
near horizon symmetries (with the $U(1)$ non-trivially fibred). One can formally
dimensionally reduce along the $U(1)$ to obtain an effective four dimensional description
in terms of a black hole with $AdS_2\times S_2$ near horizon symmetries.

After introducing an appropriate ansatz, we will calculate and analyse the entropy
function. We will apply the analysis to static black holes which turn out to have
$AdS_2\times S^3$ horizons and black rings which turn out to have $AdS_3\times S^2$
horizons. We will see that these black rings are in some sense dual to the black
holes. Interestingly, we do not need to assume the $S^{3}$ and the $AdS_{3}$ geometries
--- they follow from the entropy function analysis. We will then apply our result to Lagrangians with
real special geometry.

\subsection{Set up}
\label{sec:setup}

Before proceeding to the analysis, and to justify our ansatz,
(\ref{sauron1}-\ref{sauron3}), for the near horizon geometry, we need to establish some
notation and consider the geometry of the dimensional reduction of five dimensional black
holes and black rings to four dimensional black holes.

As previously mentioned, five dimensional black holes and black rings are characterised by
their horizon topologies which are $S^3$ and $S^2\times S^1$ respectively. Assuming no
dependence on the fifth direction we can formally dimensionally reduce their near horizon
geometry to obtain an effective four dimensional description. In the case of the $S^3$ we
can dimensionally reduce along a $U(1)$ fibre and for $S^2 \times S^1$ we can
dimensionally reduce along the $S^1$. In both cases we end up with an $S^2$ topology so
that the effective four dimensional description of both five dimensional black holes and
black rings is in terms of a four dimensional black hole.

\begin{figure}[hbtp]
  \centering
  \includegraphics[height=2.1in]{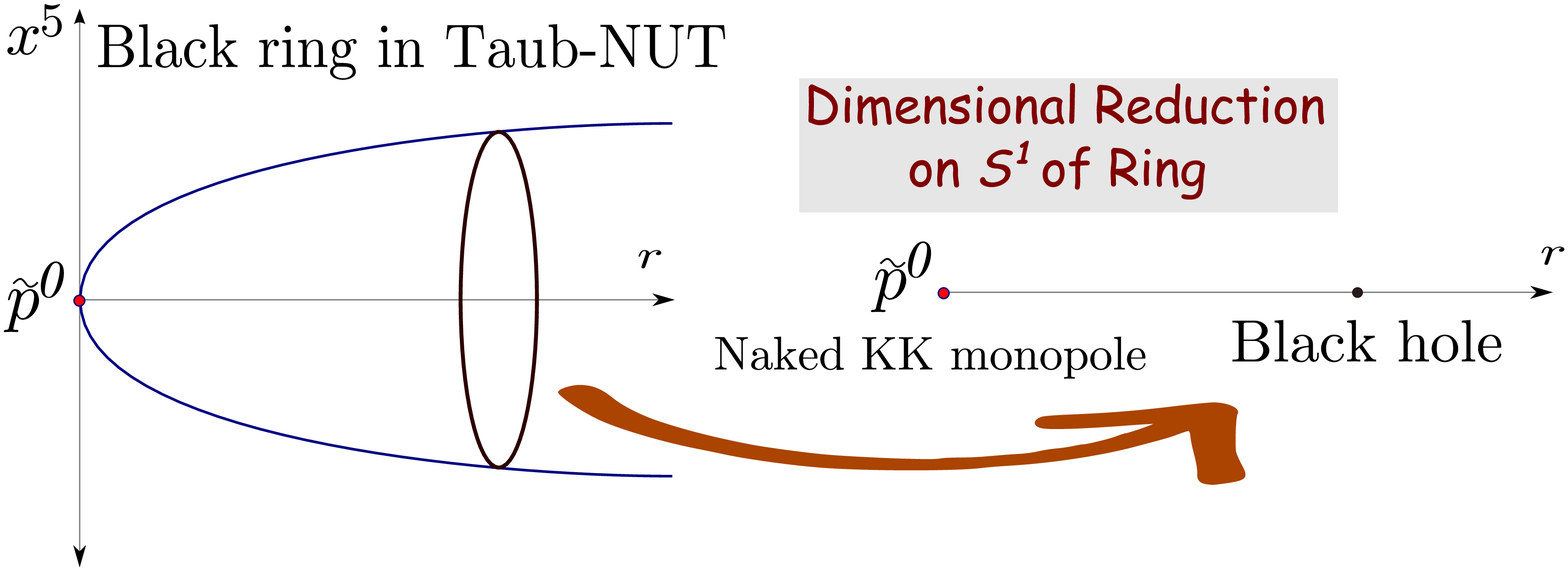}
  \caption{A Black ring away from the centre of Taub-NUT is projected down to a black hole
    and naked Kaluza-Klein magnetic monopole in four dimensions. The angular momentum carried in the
    compact dimension will translate to electric charge in four dimensions. An
    $AdS^2\times S^2 \times U(1)$ near horizon geometry will project down to
    $AdS^2\times S^2$. On the other hand, an $AdS^2 \times U(1)^2$ will go to
    $AdS^2\times U(1)$.  }
  \label{fig:ring}
\end{figure}

\begin{figure}[hbtp]
  \centering
  \includegraphics[height=2.1in]{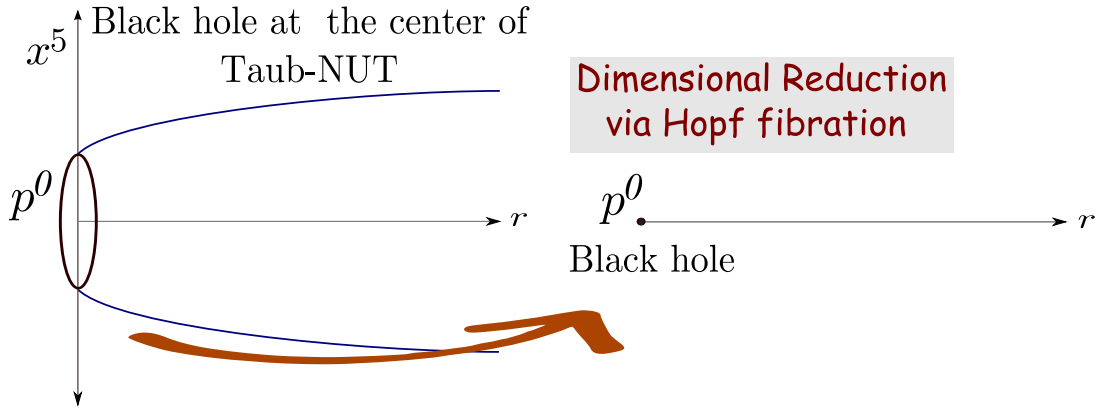}
  \hfill
  \caption{A black hole at the centre of Taub-NUT caries $NUT$ charge.  Using the Hopf
    fibration it can be projected down to black hole carrying magnetic charge. A
    spherically symmetric black hole with near horizon geometry of $AdS^2\times S^3$ will
    project down to an $AdS^2\times S^2$. On the other hand, a rotating black hole with a
    $AdS^2 \times U(1)^2$ geometry will go to $AdS^2\times U(1)$.  }
  \label{fig:bh}
\end{figure}

The dimensional reduction of black ring and black hole geometries in Taub-NUT space is
schematically illustrated in figure~\ref{fig:ring} and~\ref{fig:bh}.

\label{sec:sbh}

Since the entropy function analysis only depends on the near horizon geometry we will not
be interested in the full geometry of Taub-NUT space. We will only be concerned with its
influence on the near horizon geometry. The effect of the Taub-NUT charge is to introduce
identifications so that the black hole horizon topology becomes $S^3/{\mathbb Z}_{p^0}$
and the black ring horizon topology becomes $S^2\times S^1/ {\mathbb Z}_{{\tilde p}^0}$

We either use $\tilde{p}^0$, to denote the Taub-NUT charge of the space a black ring is
sitting in, or $p^0$, to denote the charge of a black hole sitting at the centre of the
space.  In each case the $U(1)$ will be replaced by either $U(1)/\tilde{p}^0$ or
$U(1)/{p}^0$. Unlike the black hole, the black ring does not carry Taub-NUT charge. Since
we are only looking at the near horizon geometry, the only influence of the charge on the
ring will be to induce an identification which  we can impose this by hand. To encode asymptotically flat
space we simply set the Taub-NUT charge to $1$ in both cases.  For a unified presentation,
we include ${p}^0$ and $\tilde{p}^0$ in the formulae below.  Given this notation, when we
consider black rings, we must remember to set $p^0=0$ and mod out the $U(1)$ by
$\tilde{p}^0$. When considering black holes, $p^0$ is non-zero and, since we do need to
mod out by hand, we set $\tilde{p}^0=1$.

For black holes, we can fibre the $U(1)$ over the $S^2$ to get $S^3/\mathbb{Z}_{p^0}$
while for the rings it will turn out that we can fibre the $U(1)$ over the $AdS_2$ to get
$AdS_3/\mathbb{Z}_{\tilde{p}^0}$.  These fibrations will only work for specific values of
the radius of the Kaluza-Klein circle, $w$, depending on the radii of the base spaces,
$S^2$ or $AdS^2$, and the parameters, $p^0$ or $e^0$
respectively.\footnote{$e^0$ is conjugate to the angular momentum of the ring.} Even though
we start out treating $w$ as an arbitrary parameter, we will see below that the
``correct'' value for $w$ will be dynamically generated by solving the equations of motion
for $w$ coming from the entropy function analysis. The fibration which gives us $S^3$ is
the standard Hopf fibration and the one for $AdS$, which is very similar, is discussed
towards the end of Appendix~\ref{blah}.

Now, to study the near horizon geometry of black holes and black rings in Taub-NUT space,
with the required symmetries, we specialise our Kaluza-Klein ansatz,
(\ref{sauron1b}-\ref{sauron3b}), to
\begin{align}
  ds^2&=w^{-1}\left[ v_1\left(-r^2dt^2 +{\frac{dr^2}{r^2}}\right)
    +v_2\,\left(d\theta^2+\sin^2\theta d\phi^2\right)
  \right]\nonumber\\
  &+w^2\left(d\psi+e^0\, rdt+p^0\cos\theta d\phi\right)^2,\label{sauron1}\\
  {\bar A}^I&=e^I\, rdt+{p^I\cos\theta}d\phi
  +a^I\left(d\psi+e^0\, rdt+p^0\cos\theta d\phi\right),\label{sauron2}\\
  {\bar X}^S &= u^S,\label{sauron3}
\end{align}
where the coordinates, $\theta$ and $\phi$ have periodicity $\pi$ and $2\pi$ respectively.
The coordinate $\psi$ has periodicity $4\pi$ for black holes and $4\pi/\tilde{p}^0$ for
black rings.  This ansatz, (\ref{sauron1}-\ref{sauron3}), is consistent with the near
horizon geometries of the solutions of
\cite{Elvang:2004rt,Bena:2004de,Elvang:2004ds,Gauntlett:2004qy} as discussed in Appendix~\ref{blah}.

Now that we have an appropriate five dimensional ansatz, we can construct the entropy
function from the dimensionally reduced four dimensional Lagrangian. From the four
dimensional action, we can evaluate the reduced Lagrangian, $f$, evaluated at the horizon
subject to our ansatz. The entropy function is then given by the Legendre transformation
of $f$ with respect to the electric fields and their conjugate charges.

The reduced four dimensional action, $f$, evaluated at the horizon is given by
\begin{equation}
  f= \frac{1}{16\pi G_4}\int_H d\theta d\phi  \sqrt{-g}{\cal L}
  =\frac{1}{16\pi}\left(\frac{4\pi}{\tilde{p}^0 G_5}\right)
  \int_H d\theta d\phi  \sqrt{-g}{\cal L}.
\end{equation}
The equations of motion are equivalent to
\begin{equation}\label{eom1}
  f_{,v_1}=f_{,v_2}=f_{,w}=f_{,\vec a}=f_{,\vec \Phi}=0,
\end{equation}
\begin{equation}\label{eom2}
  f_{,e^i}=N q_i,
\end{equation}
where $e^i=(e^0,e^I)$ and $q_i$ are its (conveniently normalised) conjugate charges. We
choose the the normalisation $N=4\pi/\tilde{p}^0 G_5=1/G_4$.  Using the ansatz,
(\ref{sauron1}), we find
\begin{align}
  f&= {\left(\frac{2\pi}{\tilde{p}^0 G_5}\right)} \left\{
    \begin{array}{ll}
      v_1-v_2
      &-(v_1/v_2)\left[\quater{w^3(p^0)^2}+w {\bar f}_{IJ} (p^I+p^0a^I)(p^J+p^0a^J)\right] \\
      & +(v_2/v_1)\left[\quater{w^3(e^0)^2}
        + w{\bar f}_{IJ}(e^I+e^0\,a^I)(e^J+e^0\,a^J)\right]
    \end{array}
  \right\} \nonumber \\
  & +\left(\frac{24\pi}{\tilde{p}^0 G_5}\right) {\bar c}_{IJK}\left\{ p^Ie^Ja^K+\half
    (p^0e^I+e^0p^I)a^Ja^K+\tfrac{1}{3}p^0e^0a^Ia^Ja^K \right\},
\end{align}
while (\ref{eom2}) gives the following relationship between the electric fields, $e^i$, and
their conjugate charges $q_i$:
\begin{align}
  \hat{q}_I  &=
  (v_2 / v_1)w {\bar f}_{IJ}(e^J+e^0a^J),\label{eq:q_Ihat}\\
  \hat{q}_{0} -a^I\hat q_I &= \left(v_2 / v_1\right)(\quater w^3
  e^0),\label{eq:q_0hat}
\end{align}
where, $p^i=(p^0,p^I)$, $\tilde f_{ij}$ is given by (\ref{fijt}) and the shifted charges,
$\hat{q_i}$, are defined as
\begin{align}
  \hat q_i &= q_i - \tilde f_{ij}p^j.
  \label{eq:q_ihat:deff}
\end{align} The entropy function
is the Legendre transform of $f$ with respect to the charges $q_i$:
\begin{align}
  \label{def:E}
  {\cal E}&=2\pi(N q_ie^i-f).
\end{align}
In terms of ${\cal E}$ the equations of motion become
\begin{equation}
  \label{E:eom}
  {\cal E}_{,v_1}={\cal E}_{,v_2}={\cal E}_{,w}={\cal E}_{,\vec a}={\cal E}_{,\vec\Phi}=
  {\cal E}_{,\vec e}=0.
\end{equation}
Evaluating the entropy function gives
\begin{equation}
  \label{eq:e:final}
  {\cal E}= 2\pi(Nq^ie_i-f)
  = {4\pi^2\over \tilde{p}^0 G_5} \left\{v_2-v_1
    +( v_1/v_2)
    V_{eff}
  \right\},
\end{equation}
where we have defined the effective potential
\begin{align}
  \label{eq:Veff}
  V_{eff}&= f^{ij}\hat q_i\hat q_j+f_{ij} p^i p^j
\end{align}
where $f_{ij}$ is given by (\ref{fij})
and $f^{ij}$, the inverse of $f_{ij}$, is given by
\begin{align}
  f^{ij}&=
  \bordermatrix{& 0 & J \cr 0& 4 w^{-3} & -4w^{-3} a^J \cr I& -4w^{-3} a^I & w^{-1}
    {\bar f}^{IJ} + 4w^{-3}a^Ia^J\cr }
  \label{eq:fij:inv},
\end{align}
where ${\bar f}^{IJ}$ is the inverse of ${\bar f}_{IJ}$. More explicitly, the effective
potential is given by
\begin{align}
  V_{eff} &=\quater w^3 (p^0)^2
  +4w^{-3}\{ q_0-\tilde f_{0j}(\vec a)p^j-a^I( q_I-\tilde f_{Ij}(\vec a)p^j\}^2\nonumber\\
  &+w{\bar f}_{IJ}(\vec X)\{p^I+a^Ip^0\}\{p^J+a^Jp^0\} \nonumber\\
  &+w^{-1}{\bar f}^{IJ}(\vec X) \{ q_I-\tilde f_{Ik}(\vec a)p^k\}\{ q_J-\tilde f_{Jl}(\vec
  a)p^l\},
  \label{eq:Veff:full}
\end{align}

\subsection{Charges}

From a four dimensional perspective, the charges are simple to interpret --- the $p^{i}$
are conventional magnetic charges and the $q^{i}$ are the conjugates to the electric
field. Since we are using dimensional reduction to perform our calculations, it is easiest
to work with these charges. When we write the gauge field in terms of a Kaluza-Klein
ansatz, (\ref{sauron2}), from a five dimensional perspective, things are a little more
complicated. We need to separately consider the charges $p^{0}$, $p^{I}$, $q^{0}$ and
$q^{I}$.

The charge $p^{0}$ corresponds to the Taub-NUT charge while the $p^{I}$ are related to the
dipole charge. When $p^{0}$ is zero, the charges $p^{I}$ correspond to dipole charges of the
$S_{2}$ parameterised by $\theta$ and $\phi$. This is the case for the black ring
solutions considered in section~\ref{sec:br}. On the other hand, when $p^{0}$ is non-zero,
the flux through the $S_{2}$, in our conventions, goes like $p^{I}+a^{I}p^{0}$.
It is this quantity should be interpreted as the dipole charge rather than $p^{I}$. So generally, the
relationship between $p^{I}$ and the dipole charge will depend on the value of the axions,
$a^{I}$ and the Taub-NUT charge. When we are considering black holes, we expect the dipole charge to be zero, but
as we will see in section~\ref{static}, the $p^{I}$ are non-zero. In this case they are
simply to be interpreted as a quantity proportional to $a^{I}$.

The charge $q^{0}$ is related to the angular momentum while the $q^{I}$ are related to the
electric charge. When $e^{0}$ is zero, the $q^{I}$ are simply the conjugates to the
electric-field. Analogous to the dipole charges discussed above, when $e^{0}$ is non-zero,
the electric field goes like $e^{I}+a^{I}e^{0}$. In this case, the relationship between $q^{I}$
and the electric charge depends on the values of $e^{0}$ and $a^{I}$.

\subsection{Preliminary analysis}
\label{sec:prelim}

While the effective potential $V_{eff}$ is in general quite complicated, the dependence of
the entropy function, (\ref{eq:e:final}), on the $S^2$ and $AdS^2$ radii is quite simple.
Extremising the entropy function with respect to $v_1$ and $v_2$, one finds that, at the
extremum,
\begin{equation}
  \label{eq:E:extr:2}
  {\cal E}={4\pi^2\over \tilde{p}^0 G_5} V_{eff}|_{\p V=0},
\end{equation}
with
\begin{align}
  \label{eq:v:sol}
  v_1=v_2= V_{eff}|_{\p V=0},
\end{align}
where the effective potential is to be evaluated at its extremum:
\begin{equation}
  \label{eq:Veff:extr}
  {\p_{\{w,\vec a, \vec X\}} V_{eff}}=0.
\end{equation}
From, (\ref{eq:v:sol}), we see that the radii of the $S^2$ and $AdS^2$ are equal with the
scale set by size of the charges.

As a check, we note that, the result, (\ref{eq:E:extr:2}), agrees with the
both the four and five Hawking-Bekenstein entropy since,
\begin{equation}
  \label{eq:area}
  A_{H}^{(5)}=\int d\psi d\theta d\phi
  \sqrt{g_{\psi\psi}g_{\theta\theta}g_{\phi\phi}}
  ={16\pi^{2}v_{2}}/{{\tilde p}^{0}}
\end{equation}
so
\begin{equation}
  \label{s:bh}
  {\cal E} 
  \refeqq{eq:v:sol}{eq:area}
  {A_{H}^{(5)}\over 4 G_5}
  \refeq{eq:G4vsG5} 
  {A_{H}^{(4)}\over 4 G_4}
  =
  S_{BH}.
\end{equation}
Notice that $w$ drops out of (\ref{s:bh}).

Finding extrema of the general effective potential, $V_{eff}$, given
by~(\ref{eq:Veff:full}) may in principle be possible but in practice not simple.  In the
following sections we consider simpler cases with only a subset of charges turned on.

\subsection{Black rings}
\label{sec:br}

We are now really to specialise to the case of black rings.  As discussed at the beginning
of the section, for black rings, we take $p^0=0$ so that our $AdS_2\times U(1)\times S^2$
ansatz\footnote{It will turn out that once we solve the equations of motion, the value of
  $w$ is such that the geometry is $AdS_3\times S^2$. In Appendix~\ref{blah}, we have discussed the
  near horizon geometry of supersymmetric black ring solution.} becomes
\begin{align}
  \label{sauron:ring}
  ds^2&=w^{-1}\left[ v_1\left(-r^2dt^2 +{\frac{dr^2}{r^2}}\right)
    +v_2\,\left(d\theta^2+\sin^2\theta d\phi^2\right) \right]
  +w^2\left(d\psi+e^0\, rdt\right)^2,\\
  {\bar A}^I&=e^I\, rdt+{p^I\cos\theta}d\phi+a^I\left(d\psi+e^0\, rdt\right),\label{sauron:ring2}\\
  {\bar X}^S &= u^S.\label{sauron:ring3}
\end{align}

In this case the gauge field (or in 4-D language the axion) equations simplify
considerably and it is convenient to analyse them first.  Varying $f$ with respect to
$\vec a$ we find
\begin{equation}
  \label{eq:ring:gauge}
  d_{IJ}(e^J+e^0a^J)=0
\end{equation}
where $d_{IJ}=w {\bar f}_{IJ}e^{0}+6{\bar c}_{IJK}p^K$. Assuming $d_{IJ}$ has no zero
eigenvalues,~(\ref{eq:ring:gauge}) implies that the electric field, $F^J_{tr}=e^J+e^0a^J$, is
zero. Using~(\ref{eq:q_Ihat},\ref{eq:q_ihat:deff}) this in turn implies $\hat q_I=0$
which, using~(\ref{fijt},\ref{eq:q_ihat:deff}), allows us to solve for the axions:  
\begin{equation}
    a^K= {\bar c}^{KJ}q_J,\label{eq:ring:A:sol2}
\end{equation}
where ${\bar c}^{KJ}$ is the inverse of 
\begin{equation}
  \label{eq:cij:def}
 {\bar c}_{IJ}= 6 {\bar c}_{IJK}p^K.
\end{equation}
 Notice that
$ {\bar c}_{IJ}$ is equal to  the sub-matrix, $\tilde {f}_{IJ}$, with $a^K$
replaced by $p^K$.  Now substituting (\ref{eq:ring:A:sol2}) into the definition of
$\hat{q}_{0}$ we find:
\begin{equation} 
  \label{eq:ring:A:sol}
  \hat q_0 = q_0 -\half
  {\bar c}^{IJ}q_Iq_J.
\end{equation}
So, eliminating the axions and using $\hat{q}_{I}=0$, the effective potential becomes
\begin{equation}
  V_{eff}=w{\bar f}_{IJ}p^Ip^J+(4w^{-3})(\hat q_0)^2.
\end{equation}
Using $\p_w V_{eff}=0$ we find
\begin{equation}
  \label{eq:w:ring:sol}
  w^4= {12\hat q_0^2\over V_{M}}
\end{equation}
where we have defined the magnetic potential, $V_{M}={\bar f}_{IJ}p^Ip^J$. So
\begin{equation}
  V_{eff}={4\over 3}w V_M = 16 w^{-3}(\hat q_0)^2\label{eq:VM:extr}
\end{equation}
Eliminating $w$ from $V_{eff}$ we get 
\begin{equation}
  {\cal E}
  ={8\pi^2\over \tilde{p}^0 G_5} \sqrt{
    \hat q_0
    (\tfrac{4}{3} V_{M})^{3\over2}}.\label{eq:E:VM:extr}
\end{equation}
We note that
\begin{equation}
  \label{eq:ads3:embedding}
  e_0^2w^2=v_1w^{-1}
\end{equation}
which, we see by comparison with (\ref{eq:final}), means that we have a
$S^2\times AdS_3/\mathbb{Z}_{\tilde{p}^0}$ near horizon geometry. Finally,
using~(\ref{eq:v:sol},\ref{eq:VM:extr},\ref{eq:ads3:embedding}) we can also write the
entropy as
\begin{equation}
  \label{eq:E:alt}
  {\cal E}={4\pi^2\over \tilde{p}^0 G_5}{(\tfrac{4}{3} V_{M})^{3\over2}}{(e^{0})^{-1}}
\end{equation}

\subsection{Static 5-d black holes}
\label{static}

We now consider five dimensional static spherically symmetric black holes. Since they are
not rotating we take $e^0=0$. This is in some sense ``dual'' to taking $p^0=0$ for black
rings. To examine this analogy further, we will relax the natural assumption of an
$AdS_2\times S^3$ geometry to $AdS_2\times S^2\times U(1)$. We will see that the analysis
for the black holes is very similar to the analysis of the black rings with the magnetic
potential replaced by an electric potential.  Once we solve the equations of motion we
recover an $AdS_2\times S^3$ geometry via the Hopf fibration.  This is analogous to the
black ring where we got $AdS_3\times S^2$ with the $U(1)$ fibred over the $AdS_2$ rather
than the $S^2$.

With $e^0=0$, our ansatz becomes
\begin{align}
  \label{sauron1c}
  ds^2&=w^{-1}\left[ v_1\left(-r^2dt^2 +{\frac{dr^2}{r^2}}\right)
    +v_2\,\left(d\theta^2+\sin^2\theta d\phi^2\right) \right]
  +w^2\left(d\psi+p^0\cos\theta d\phi\right)^2,\\
  A^I&=e^I\, rdt+{p^I\cos\theta}d\phi+a^I\left(d\psi+p^0\cos\theta d\phi\right),\label{sauron2c}\\
  \Phi^S &= u^S.\label{sauron3c}
\end{align}
In this case the gauge field equation becomes
\begin{equation}
  \label{eq:bh:gauge}
  \tilde d_{IJ}(p^J+p^0a^J)=0
\end{equation}
where $\tilde d_{IJ}=w {\bar f}_{IJ}p^{0}-6{\bar c}_{IJK}e^K$. Assuming $\tilde d_{IJ}$ has no
zero eigenvalues, (\ref{eq:bh:gauge}) implies $F^{I}_{\theta\phi}=0$,
 which, together with (\ref{eq:q_0hat},\ref{eq:q_ihat:deff}), gives
\begin{align}
  \label{eq:ring:B:sol}
  \hat q_0-a^I\hat q_i &= 0 \\
  a^K&= -p^K/p^0,\\
  \hat q_I &= q_I + 3 {\bar c}_{IJK}p^Jp^K/p^0
\end{align}
and the effective potential becomes
\begin{equation}
  \label{eq:Veff:bh}
  V_{eff}=(\quater w^3)( p_0)^2+w^{-1}{\bar f}^{IJ}\hat q_I\hat q_J.
\end{equation}
Using $\p_w V_{eff}=0$ we find
\begin{equation}
  \label{eq:w:bh:sol}
  w^4= {4 V_{E}\over 3 p_0^2 }
\end{equation}
where we have defined the electric potential $V_E=\bar{f}^{IJ}\hat q_I\hat q_J$.  So
\begin{equation}
  V_{eff}={4\over 3}w^{-1} V_E=w^3(p^0)^2
\end{equation}
Eliminating $w$ from ${\cal E}$ we find
\begin{equation}
  {\cal E}
  ={4\pi^2\over G_5}
  \sqrt{
    p_0
    (\tfrac{4}{3} V_{E})^{3\over2}
  }.
\end{equation}
We note that, analogous to the ring case where we had $e_0^2w^2=v_1w^{-1}$,
\begin{equation}
  \label{eq:s3:embedding}
  p_0^2w^2=v_2w^{-1}
\end{equation}
which, via the Hopf fibration, gives us an $AdS_2\times S^3/\mathbb{Z}_{p^0}$ near horizon
geometry.

\subsection{Very Special Geometry}
\label{sec:mag}

We now consider the explicit example of ${\cal N}=2$ supergravity in five dimensions
corresponding to M-theory on a Calabi-Yau threefold -- this gives what has been called
real or very special geometry
\cite{Chamseddine:1996pi,Gunaydin:1983bi,Gunaydin:1984ak,deWit:1992cr,9506144,Papadopoulos:1995da,Antoniadis:1995vz}. Some
properties of very special geometry which we use are recorded in
Appendix~\ref{sec:special}.  Building on the general results of the previous sections, to
find the attractor values of the scalars and the entropy we just need to extremise the
relevant magnetic or electric potentials.

\subsubsection{Black rings and very special geometry}
\label{sec:mc}

For  very special geometry, the magnetic potential is given by
\begin{equation}
  V_M = \bar{f}_{IJ}p^{I}p^{J}\refeq{eq:hvsf}\tfrac{1}{2} H_{IJ}p^Ip^J
\end{equation}
where the properties of $H_{IJ}$ can be found in  Appendix~\ref{sec:special}.

Extremising the magnetic potential gives
\begin{equation}
  \label{eq:extr:M}
  \p_i V_{M} = \tfrac{1}{2}\p_i(H_{IJ}p^Ip^J) \refeq{special:lower:2} \tfrac{1}{4} \p_i(p_Ip^I)=0
\end{equation}
These equations have a solution
\begin{equation}
  \label{BPS:anz}
  \lambda X_I= p_I
\end{equation}
This condition follows from one of the BPS conditions found in \cite{Kraus:2005gh}. To see
that (\ref{BPS:anz}) is indeed a solution, we insert it into
(\ref{eq:extr:M}), which gives
\begin{eqnarray}
  \p_i(p_Ip^I) &=& \p_i(p_I)p^I\\
  &\refeq{BPS:anz}&\lambda\p_i(X_I)p^I\label{kl}\\
  &\refeq{special:covariant}&-2\lambda H_{IJ}p^I\p_i(X^J)\\
  &\refeq{special:lower}&-\lambda p_J\p_i(X^J)\\
  &\refeq{BPS:anz}&-\lambda^2 X_J\p_i(X^J)\refeq{X:norm:2}0
\end{eqnarray}
We can fix the constant $\lambda$ using (\ref{vol}) which gives
\begin{equation}
  X_I= \frac{p_I}{(\frac{1}{6}C_{IJK}p^Ip^Jp^K)^{\frac{1}{3}}}\label{eq:ring:scalar:attr}
\end{equation}
so finally we get for $X^I$,
\begin{equation}
  X^I=
  \frac{p^I}{({1 \over 6}C_{IJK}p^Ip^Jp^K)^{\frac{1}{3}}}
\end{equation}
and
\begin{align}
  V_{M}|_{\p V=0} &=\tfrac{1}{2}H_{IJ}p^Ip^J
  =\tfrac{1}{2}\lambda^2 H_{IJ}X^IX^J={3 \over 4}\lambda^2\\
  &=\frac{3 }{4} \left( \frac{1}{6}C_{IJK}p^Ip^Jp^K \right)^{\frac{2}{3}}\label{eq:vm:extr}
\end{align}
This is the supersymmetric solution of \cite{Kraus:2005gh} derived from the BPS attractor
equations.

Notice that (\ref{kl}) can be rewritten as extremising the magnetic central charge,
$Z_M=X_Ip^I$:
\begin{equation}
  \p_i(X_I)p^I= \p_iZ_M=0
\end{equation}
So we see that $\p_iV_M=0$ together with the BPS condition (\ref{BPS:anz}) implies $Z_M$
extremised. The converse is not necessarily true suggesting there are non-BPS black ring
extrema of $V_M$ --- this is discussed below.

Now, from (\ref{eq:vm:extr},\ref{eq:E:VM:extr}) we find that the entropy is 
\begin{equation}
  \label{eq:E:special}
  {\cal E}={8\pi^2\over \tilde{p}^0 G_5} \sqrt{
    {\hat q_0}(\tfrac{1}{6} C_{IJK}p^{I}p^{J}p^{K})
    } ={4\pi^2\over \tilde{p}^0 G_5}{(\tfrac{1}{6} C_{IJK}p^{I}p^{J}p^{K})}{(e^{0})^{-1}}
\end{equation}
As discussed in Appendix~\ref{blah}  this gives the correct entropy 
for the special case of the ring solution of~\cite{Elvang:2004ds}.

\subsubsection{Static black holes and very special geometry}
\label{sec:elec}

The analysis for these black holes is analogous to the black rings. From the attractor
equations for a static black hole, governed by
\begin{equation}
  \label{eq:ve:special}
  V_E=\bar{f}^{IJ}\hat q_I\hat q_J=2 H^{IJ}\hat q_I\hat q_J
\end{equation}
 we will get the equation:
\begin{equation}
  \label{eq:extr}
  \p_i V_{E} = 2\p_i(H^{IJ}\hat q_I\hat q_J)=  \p_i(\hat q^I\hat q_I)=0
\end{equation}
This will have similar solutions
\begin{equation}
  X^I= \frac{\hat q^I}{({1 \over 6}C_{IJK}\hat q^I\hat q^J\hat q^K)^{\frac{1}{3}}}
\end{equation}
Similarly, extremising the electric central charge $Z_e$ of \cite{Kraus:2005gh} together
with the BPS condition implies $V_E$ is extremised. The converse is not necessarily true
suggesting there are non-BPS black hole extrema of $V_E$ as noted in \cite{Larsen:2006xm}.

In a similar fashion to the black ring case, we find that the entropy is given by
\begin{equation}
  \label{eq:E:special:2}
  {\cal E}=\frac{\pi^2}{2 G_5} \sqrt{
    {p^0}(\tfrac{1}{6} C^{IJK}[16{\hat q}_{I}][16{\hat q}_{J}][16{\hat q}_{K}])
    }.
\end{equation}
which, modulo a different normalisation for the charges, is the same as the entropy quoted
in \cite{Larsen:2006xm} (albeit modified due to the presence of a Taub-NUT charge). As
shown in Appendix~\ref{sec:spherical:nh}, our charges, $\hat{q}_{I}$, are related to those
of \cite{Larsen:2006xm} by
\begin{equation}\label{eq:diff:norm}
  {16\hat q}_{I}=Q_{I}.
\end{equation}
The appearance of the shifted charge ${\hat q}_{I}$  rather than $q_{I}$ is due to the
Chern-Simons term.

\subsubsection{Non-supersymmetric solutions of very special geometry}

In 4 dimensional ${\cal N}=2$ special geometry we can write $V_{eff}$ or the ``blackhole
potential function'' as \cite{Kallosh:2005ax}
\begin{equation}
  V_{BH}= |Z|^2 +|DZ|^2.
\end{equation}
As noted in \cite{Kallosh:2005ax} and \cite{Goldstein:2005hq} (in slightly different
notation), for BPS solutions, each term of the potential is separately extremised while
for non-BPS solutions $V_{BH}$ is extremised but $DZ\neq0$. It is perhaps not surprising
that a similar thing happens in very special geometry.
In fact, this generalisation of the non-BPS attractor equations to five dimensional static
black holes has already be shown in \cite{Larsen:2006xm} using a reduced Lagrangian
approach.

The electric potential $V_E$ can be written
\begin{equation}
  \tfrac{1}{2}V_E=H^{IJ}\hat q_J\hat q_J= H^{IJ}(D_{I}\hat Z_{E})(D_{J}\hat Z_{E})+\tfrac{2}{3}(\hat Z_{E})^{2}.
\end{equation}
Solving $D_IV_E=0$ we find a BPS solution, $D_I\hat Z_E=0$, and another solution
\begin{equation}
  \tfrac{2}{3}H_{IJ}\hat Z_{E}+D_{I}D_{J}\hat Z_{E}=0.
\end{equation}
Similarly, we find the magnetic potential, $V_M$, can be written
\begin{equation}
  2V_{M}=H_{IJ}p^Ip^J=\tfrac{1}{3}Z_{M}^{2}+H^{IJ}D_IZ_{M}D_JZ_{M}
\end{equation}
and solving $D_IV_M=0$ we find a BPS solution, $D_I Z_M=0$, and another solution
\begin{equation}
  \tfrac{1}{3}H_{IJ}Z_{M}+D_{I}D_{J}Z_{M}=0.
\end{equation}

We conjecture one can obtain some five dimensional non-SUSY solutions by lifting non-SUSY
solutions in four dimensions which have $AdS_2\times S^2$ near horizon geometries using
the $4D$-$5D$ lift. Furthermore the analysis of \cite{Goldstein:2005hq} should go through
so that for such solutions to exist we require that extremum of $V_{eff}$ is a minimum --
in other words the matrix
\begin{equation}
  \left.{\p^2V_{eff}\over\p\Phi^S\p\Phi^T}\right|_{\p V=0}>0
\end{equation}
should have non-zero eigenvalues.

\sectiono{General Entropy function}
\label{sec:gen}

We now relax our symmetry assumptions to $AdS_2\times U(1)^2$, taking the following ansatz
\begin{align}
  \label{eq:anz:ef}
  ds^2&=w^{-1}(\theta)\Omega^2(\theta)e^{2\Psi(\theta)}
  \left(-r^2dt^2+{dr^2\over r^2}+\beta^2d\theta^2 \right)
  +w^{-1}(\theta)e^{-2\Psi(\theta)}(d\phi+e_\phi r dt)^2 \nn\\
  &+w^2(\theta)(d\psi+e_0 rdt+b_0(\theta)d\phi)^2\\
  A^I &= e^I r d t + b^I(\theta)(d\phi+e_\phi r dt)
  + a^I(\theta)(d\psi+e_0 rdt+b_0(\theta)d\phi)\\
  \phi^S &= u^S(\theta).
\end{align}
Now, using (\ref{eq:4d_action}) and then following \cite{Astefanesei:2006dd}, the entropy
function is
\begin{align}
  {\cal E} &\equiv 2\pi (J_\phi e_\phi + \vec q \cdot \vec e - \int d\theta d\phi
  \sqrt{-\det g} \, {\cal L})
  \\
  &= 2\pi (J_\phi e_\phi +  \vec q \cdot \vec e) \nonumber\\
  &- {\pi^2\over\tilde{p}_0G_5} \int d\theta \, \Bigg[2\Omega^{-1} \beta^{-1} (\Omega')^2
  - 2\Omega \beta - 2 \Omega \beta^{-1} (\Psi')^2 +{1\over 2} \alpha^2 \Omega^{-1} \beta
  e^{-4\Psi} -\beta^{-1}\Omega h_{rs}(\vec u)
  u_r' u_s' \nonumber \\
  &+ 4 \tilde f_{ij} (\vec u) (e_i -\alpha b_i) b_j' + 2 f_{ij}(\vec u)
  \left\{\beta\Omega^{-1} e^{-2\Psi} (e_i - \alpha b_i) (e_j - \alpha b_j) -
    \beta^{-1}\Omega e^{2\Psi}b_i' b_j' \right\}
  \Bigg] \nonumber \\
  & +{2\pi^2\over\tilde{p}_0G_5}\, \left[ \Omega^2 e^{2\Psi} \sin\theta (\Psi' + 2\Omega'
    / \Omega )\right]_{\theta=0}^{\theta=\pi}.
\end{align}
where $f_{ij}$, $\tilde f_{ij}$,$h_{rs}$ and $u^s$ related to five dimensional quantities
as discussed in section~\ref{sec:thing}. Now extremising the entropy function gives us
differential equations.

Using the near horizon geometry of the non-SUSY black ring of \cite{Elvang:2004xi}, which
we evaluate in Appendix~\ref{sec:nonsusy}, we find that the entropy function gives the
correct entropy.

\bigskip

{\bf Acknowledgements: } We would like to thank Sandip Trivedi for helpful discussions and
invaluable comments on the draft and Ashoke Sen for useful comments and suggestions. We
thank the organisers of the conference ``ISM'06'', held in Dec 2006, where some part of
the work was done. We also thank the people of India for supporting research in String
theory.

\appendix

\section{Dimensional reduction}\label{sec:dr}

\newcommand{\I}{\left(\int dy\right)\int d x^4 \sqrt{- g}\; }
\newcommand{\cijkb}{\; \bar{c}_{IJK}}
\newcommand{\cijk}{\bar{c}_{IJK}}

\newcommand{\et}{\tilde\epsilon^{\mu\nu\alpha\beta}}
\newcommand{\etb}{{\tilde\epsilon}^{\bar\mu\bar\nu\bar\alpha\bar\beta\bar\delta}}

In this section present some details of the dimensional reduction of the Chern-Simons
term. We start with the five dimensional
gauge-fields which we assume are independent of the fifth direction:
\begin{eqnarray}
  \bar{A}^{I}&=&\bar{A}^{I}_{\mu}dx^{\mu}+\bar{A}^{I}_{y}dy \\
 &=&A^I_\mu dx^\mu+a^I(x^\mu)\left(dy+A^0_\mu dx^\mu\right).
\end{eqnarray}
From these definitions we can relate the five dimensional gauge field strength  to four dimensional quantities as follows
\begin{align}
\bar{F}_{\mu\nu}^{I}&=F_{\mu\nu}^{I}+a^{I}
F_{\mu\nu}^{0}+(\partial_{\mu}a^{I})A_{\nu}^{0}-(\partial_{\nu}a^{I})A_{\mu}^{0},  \\
\bar{F}_{\nu y}^{I}&=\partial_{\nu}a^{I},  \label{eq:fmuy}
\end{align}
where 
\begin{equation} 
F_{\mu\nu}^{i}=\partial_{\mu}A_{\nu}^{i}-\partial_{\nu}A_{\mu}^{i}\qquad i=I\mbox{ or }0.
\end{equation}
We can now write a five dimensional Chern-Simons term,
\begin{equation}
  \label{eq:lc}
  \sqrt{-\bar g}{\cal L}_{CS} = \cijkb
  \etb
  \bar{A}^{I}_{\bar \mu} 
  \bar{F}^{J}_{\bar\alpha\bar\beta}
  \bar{F}^{K}_{\bar\mu\bar\nu},
\end{equation}
 as
\begin{align}
  \label{eq:cs1}
  \sqrt{-\bar g}{\cal L}_{CS}&=
  \cijk\et(\bar{A}^{I}_{y}\bar{F}^{J}_{\mu\nu}\bar{F}^{K}_{\alpha\beta}
  +4\bar A^{I}_{\mu}\bar{F}^{J}_{\nu y}\bar{F}^{K}_{\alpha\beta})\\
  &{\rightarrow}\;
  3\cijk\et a^{I}\bar{F}^{J}_{\mu\nu}\bar{F}^{K}_{\alpha\beta}\\
  &=
 3\cijk\et 
  \left(
    a^{I}(F_{\mu\nu}^{J}+a^{J} F_{\mu\nu}^{0})
  (F_{\alpha\beta}^{K}+a^{K} F_{\alpha\beta}^{0})
   +
    \cancel{4a^{J}_{,\mu}a^{K}_{,\alpha}A_{\nu}^{0}A_{\beta}^{0}}
  \right)\nonumber\\
 & +3\cijk\et 
    \Big(
    \underbrace{
      \tfrac{4}{2} (a^{I} a^{J})_{,\mu}A_{\nu}^{0}F_{\alpha\beta}^{K}
    }_{
      \rightarrow -a^{I} a^{J}F_{\mu\nu}^{0} F_{\alpha\beta}^{K}
    } 
    +\underbrace{ 
      \tfrac{4}{3} (a^{I} a^{J} a^{K})_{,\mu}A_{\nu}^{0}F_{\alpha\beta}^{0}
    }_{
      \rightarrow-\tfrac{2}{3} a^{I}a^{J}a^{K}F_{\mu\nu}^{0}F_{\alpha\beta}^{0}
    }  
    \Big)\\
 &= 3\cijk\et  \left(
 a^{I}F_{\mu\nu}^{J}F_{\alpha\beta}^{K}
 +\tfrac{1}{2}a^{I}a^{J}(F_{\mu\nu}^{K}F_{\alpha\beta}^{0}+F_{\mu\nu}^{0}F_{\alpha\beta}^{K})
 + \tfrac{1}{3}a^{I}a^{J}a^{K}F_{\mu\nu}^{0}F_{\alpha\beta}^{0} 
 \right)\\
&=  (\tfrac{1}{2}\tilde{f}_{ij}) F^{i}_{\mu\nu}F^{j}_{\alpha\beta}\et 
\end{align}
where the arrow, ``$\rightarrow$'', denotes the use of integration by parts, and
$\tilde\epsilon^{01234}=\tilde\epsilon^{0123}=1$, are the completely antisymmetric
Levi-Civita symbols.

\section{Notes on Very Special Geometry}
\label{sec:special}
\setcounter{equation}{0}

Here we collect some useful relations and define some notation from very special geometry
along the lines of \cite{Kraus:2005gh,Larsen:2006xm}, which are used in section~\ref{sec:mag}.
\begin{itemize}
\item We take our CY$_3$ to have Hodge numbers $h^{1,1}$ with the index
  $I\in{1,2, \ldots , h^{1,1}}$.
\item The K\"{a}hler moduli, $X^I$ which are {\it real}, correspond to the volumes of the
  2-cycles.
\item $C_{IJK}$ are the triple intersection numbers. They are related to the couplings
  defined in (\ref{acthd}) by
  \begin{equation}
    \label{eq:c_vs_bar_c}
    C_{IJK}={4!}\bar{c}_{IJK}
  \end{equation}
\item The volumes of the 4-cycles $\Omega_I$ are given by
  \begin{equation}
    \label{XI}
    X_I  = {1\over 2} C_{IJK} X^J X^K~.
  \end{equation}
\item The prepotential is given by
  \begin{equation}
    \label{vol}
    {\cal V}   = {1\over
      6} C_{IJK} X^I X^J X^K =1
  \end{equation}
\item The volume constraint (\ref{vol}) implies there are $n_v= h^{1,1}-1$ independent
  vector-multiplets.
\item denote the independent vector-multiplet scalars as $\phi^i$, and the corresponding
  derivatives $\partial_i = {\p \over \p \phi^i}$.
\item The kinetic terms for the gauge fields are governed by the metric
  \begin{equation}
    \label{special:met}
    {H_{IJ}  =
      -\half \left.\p_I \p_J \ln {\cal V}\right|_{{\cal V}=1} =-\half( C_{IJK} X^K -  X_I
      X_J)~,}
  \end{equation}
  where we use the notation for derivatives: $\p_I = {\p \over \p X^I}$. In terms of the
  couplings used in (\ref{acthd}) we have
  \begin{equation}
    H_{IJ}={\bar h}_{IJ}=2\bar{f}_{IJ}\label{eq:hvsf}
  \end{equation}

\item The electric central charge is given by
  \begin{equation}\label{ze}
    Z_E = X^I q_I.
  \end{equation}
  We generalise this to
  \begin{equation}
    \label{eq:zehat}
    \hat Z_E = X^I \hat q_I.
  \end{equation}
\item The magnetic central charge is given by
  \begin{equation}\label{zehat}
    Z_M = X_I p^I.
  \end{equation}

\item From (\ref{XI}) it follows that
  \begin{equation} {X_I X^I =3~,}
    \label{X:norm}
  \end{equation}
  so
  \begin{equation}
    {X^I \p_i X_I  = \p_i X^I X_I =0~.}
    \label{X:norm:2}
  \end{equation}
  which in turn together with (\ref{special:met}) gives
  \begin{align}
    \quad X_I &= 2 H_{IJ}X^J~, \label{special:lower}\\
    \p_i X_I &= - 2 H_{IJ} \p_i X^J~. \label{special:covariant}
  \end{align}
\item As suggested by, (\ref{special:lower}), we will use $2 H_{IJ}$ to lower indices, so
  for example,
  \begin{equation}
    p_I = 2H_{IJ}p^J\label{special:lower:2},
  \end{equation}
  which in turn implies we should raise indices with $\half H^{IJ}$,
  \begin{equation}
    \label{special:raise}
    q^I = \half H^{IJ} q_J
  \end{equation}
  where $H^{IJ}$ is the inverse of $H_{IJ}$.
\item In order to take the volume constraint (\ref{vol}) into account, it is convenient to
  define a covariant derivative $D_I$,
  \begin{equation}
    \label{eq:DI}
    D_I f = (\p_I-\tfrac{1}{3}(\p_I\ln{\cal V})|_{{\cal V}=1})f.
  \end{equation}
  Rather than extremise with respect to the real degrees of freedom using $\p_i$, we can
  take covariant derivatives.
\end{itemize}

\section{Supersymmetric black ring near horizon geometry}\label{blah}
\setcounter{equation}{0}
Here, we will consider the black ring solution of \cite{Elvang:2004rt}, and find the near
horizon limit of the metric and the gauge fields. This will enable us to compare with the
charges defined in section~\ref{sec:setup}. 

As \cite{Elvang:2004rt} follows the
conventions of \cite{Gauntlett:2002nw} the relevant Lagrangian is
\begin{equation}
  \label{eq:ring:lag}
  {\cal L}=R-\tilde{F}_{\mu\nu}\tilde{F}^{\mu\nu}
  -\tfrac{2}{3\sqrt{3}}\tilde{F}_{\alpha\beta}\tilde{F}_{\mu\nu}\tilde{A}_{\gamma}\epsilon^{\alpha\beta\mu\nu\gamma}.
\end{equation}
We can obtain this action from very special geometry by taking, $n_{v}=3$ with  the
gauge fields equal to each other, $F^{I}_{\mu\nu}=F_{\mu\nu}$, fixing the scalars at their attractor value
(\ref{eq:ring:scalar:attr}), and taking
\begin{equation}
  \label{eq:ring:cijk}
  C_{IJK}=|\epsilon_{IJK}|
\end{equation}
where $\epsilon_{IJK}$ is the Levi-Civita symbol. This gives
\begin{equation}
  X^{I} = 1, \qquad  H_{IJ} = (\tfrac{1}{2},\tfrac{1}{2},\tfrac{1}{2}),
\end{equation}
and the Lagrangian becomes
\begin{equation}
  \label{eq:other}
  {\cal L}=R-\tfrac{3}{4}{F}_{\mu\nu}{F}^{\mu\nu}
  -\tfrac{1}{4}{F}_{\alpha\beta}{F}_{\mu\nu}{A}_{\gamma}\epsilon^{\alpha\beta\mu\nu\gamma}.
\end{equation}
Comparing (\ref{eq:ring:lag}) and (\ref{eq:other}) we find
\begin{equation}
  \label{eq:comp}
  \tilde{A}_{\mu}=\tfrac{\sqrt{3}}{2}{A}_{\mu}
\end{equation}

Now, the metric for the black ring solution of \cite{Elvang:2004rt} is
\begin{equation}
  ds^2=-f^2(dt+\omega)^2+f^{-1}ds^2(M_{4})\,,\label{metric}
\end{equation}
where
\begin{equation}
  f^{-1}=1+\frac{Q-q^2}{2R^2}(x-y)-\frac{q^2}{4R^2}(x^2-y^2)\label{simplef}
\end{equation}
\begin{align}
  ds^2(\mathbb{R}^{4})= \frac{R^2}{(x-y)^2} \left\{
    \frac{dy^2}{y^2-1}+(y^2-1)d\vartheta^2+\frac{dx^2}{1-x^2}+(1-x^2)d\phi^2\right\}
  \label{base}
\end{align}
and $\omega=\omega_{\vartheta}(x,y)d\vartheta+\omega_{\phi}(x,y)d\phi$ with
\begin{align}
  \omega_{\phi} &= -\frac{q}{8R^2}(1-x^2)\left[3Q-q^2(3+x+y)\right]\,,\label{omegas}\\
  \omega_{\vartheta} &= \frac{3}{2}q(1+y)+\frac{q}{8R^2}(1-y^2)\left[3Q-q^2(3+x+y)\right]\,.
\end{align}
The variables $\vartheta$ and $\phi$ have period $2\pi$, while $-1\leq x\leq 1$ and $\infty<y\leq-1$. 
The gauge field is expressed as,
\begin{equation}
  \tilde{A}=\frac{\sqrt{3}}{2}
  \left[f\,(dt+\omega)-\frac{q}{2}((1+x)\, d\phi+(1+y)\, d\vartheta)\right]\,.\label{apot}
\end{equation}

The ADM charges are given by
\begin{align}
  M &= \frac{3\pi}{4G}Q\,,\qquad J_{\phi}=\frac{\pi}{8G}\, q\,(3Q-q^2)\,,\nonumber \\
  J_{\vartheta} &= \frac{\pi}{8G}\, q\,(6R^2+3Q-q^2)\,.\label{adm}\end{align}

\subsubsection*{Near Horizon Geometry}
\label{sec:nh}
In these coordinates, the horizon lies at $y\rightarrow-\infty$.  To examine the near
horizon geometry, it is convenient to define a new coordinate $r=-R/y$ (so the horizon is
at $r=0$). Then consider a coordinate transformation of the form
%
\begin{align}
  dt &= dv-B(r)dr,\nonumber \\
  d\phi &= d\phi'-C(r)dr,\\
  d\vartheta &= d\vartheta'-C(r)dr,
\end{align}
%
where
%
\begin{equation}
  B(r)=\frac{B_2}{r^2}+\frac{B_{1}}{r}+B_{0},\qquad C(r)=\frac{C_{1}}{r}+C_{0}.
\end{equation}
%
where $B_2=q^2L/(4R)$ and $C_{1}=-q/(2L)$, with
%
\begin{equation}
  L\equiv\sqrt{3\left[\frac{(Q-q^2)^2}{4q^2}-R^2\right]},
\end{equation}
and
\[
B_{1}=(Q+2q^2)/(4L)+L(Q-q^2)/(3R^2)
\]
\[
C_{0}=-(Q-q^2)^{3}/(8q^{3}RL^{3})
\]
\[
B_{0}=q^2L/(8R^{3})+2L/(3R)-R/(2L)+3R^{3}/(2L^{3})+3(Q-q^2)^{3}/(16q^2RL^{3})
\]
The metric (\ref{base}) becomes
%
\begin{align}
  \label{met}
  ds^2 &= -\frac{16r^{4}}{q^{4}}dv^2
  +\frac{2R}{L}dvdr+\frac{4r^{3}\sin^2\theta}{Rq}dvd\phi'
  + \frac{4Rr}{q}dvd\vartheta'+\frac{3qr\sin^2\theta}{L}drd\phi'\nonumber \\
  &+ 2\Big[ \frac{qL}{2R}\cos\theta+\frac{3qR}{2L}+\frac{(Q-q^2)(3R^2-2L^2)}{3qRL}
  \Big]drd\vartheta'\nonumber \\
  &+ L^2d{\vartheta'}^2
  +\frac{q^2}{4}\left[d\theta^2+\sin^2\theta\left(d\phi'-d\vartheta'\right)^2\right] +\ldots
\end{align}
%
where we have neglected terms which will disappear when we take the near horizon limit:
\begin{equation}
  \label{eq:nhl}
  r=\epsilon L\tilde{r}/R, \;
  v=\tilde{v}/\epsilon,\; \epsilon\rightarrow0.
\end{equation}
The gauge field (\ref{apot}) becomes:
%
\begin{align}
  \label{eq:gauge2}
  \tilde{A}&=\half{\sqrt{3}}\Big[f\,(dv+\omega')
  -\half{q}\left(\{1+x\}\, d\phi'+\{1+y\}\,d\vartheta'\right) \\
  & -\left(fB+C\left\{
      f\omega_{\phi}+f\omega_{\vartheta}-\half{q}(1+x)-\half{q}(1-{R}/{r})\right\}
  \right)dr\Big]
\end{align}
with $\omega'=\omega_{\vartheta}d\vartheta'+\omega_{\phi}d\phi'$ In the limit of small $r$
%
\begin{align}
  \label{eq:f2}
  f&=\frac{1}{1+x(f_1-f_2 x)+f_1 r^{-1}+f_2
    r^{-2}}\\
  &=\frac{r^2}{f_2} \left( 1-f_1f_2^{-1}r+
    \left(f_1^2f_2^{-1}-1+x(x-f_1f_2^{-1})\right)f_2^{-1}r^2+\mathcal{O}\left(r^{3}\right)
  \right)
\end{align}
%
where $f_1=(Q-q^2)/2R^2$ and $f_2=q^2/4R^2$. Expanding $\omega$ in the limit of small $r$,
we have,
%
\begin{align}
  \omega_{\phi} &= \left\{ -\frac{q^{3}\left(1-x^2\right)}{8R} \right\} \frac{1}{r}
  +\left\{ \frac{q\left(xq^2+3q^2-3Q\right)\left(1-x^2\right)}{8R^2}
  \right\} \label{omega_r}\\
  \omega_{\vartheta} &= \left\{ -\frac{q^{3}R}{8}\right\} \frac{1}{r^{3}} +\left\{
    \frac{xq^{3}}{8}+\frac{3q^{3}}{8}-\frac{3Qq}{8} \right\} \frac{1}{r^2} +\left\{
    \frac{q^{3}}{8R}-\frac{3qR}{2}
  \right\} \frac{1}{r}\nn\\
  &+\left\{ -\frac{xq^{3}}{8R^2}-\frac{3q^{3}}{8R^2}+\frac{3Qq}{8R^2}+\frac{3q}2 \right\}
\end{align}
%
Expanding out the gauge field (neglecting some terms which can be gauged away) we obtain:
\begin{eqnarray}
  \tilde{A} &=
  \half{\sqrt{3}}
  \left\{ -
    \left[
      {q\over2 }+{Q\over 2q}+\mathcal{O}(r)
    \right]d\vartheta'
    +\left[-\frac{q}{2}\left(x+1\right)+\mathcal{O}(r)\right]d\chi
  \right.\nn\\
  &   +
  \left.
    \left[
      \frac{4}{q^{2}}r^{2}+\mathcal{O}(r^{3})
    \right]dv
    +\left[
      c_{r}x\, r+\mathcal{O}(r^{2})
    \right]dr
  \right\}
\end{eqnarray}
where $\chi=\phi-\vartheta$ and
\begin{equation}
  c_{r}
  =
  \frac{L\left(R^{2}\left(2R^{4}-3\right)q^{4}
      +\left(-4QR^{6}+6QR^{2}+2\right)q^{2}+Q^{2}R^{2}\left(2R^{4}-3\right)\right)}
  {2q^{2}}
\end{equation}
Finally taking the near-horizon limit (\ref{eq:nhl}), letting $x=-\cos\theta$,
$\chi\rightarrow\phi$ and $\vartheta'\rightarrow \psi/2$, we
obtain\footnote{In our conventions the third angle, $\psi$, has period $4\pi$.}
%
\begin{equation}
  \label{A:nh}
  \tilde{A}= -\frac{\sqrt{3}}{8}
  \left[
    {q}+{Q\over q}
  \right]
  d\psi
  +\frac{\sqrt{3}q}{4}\left(\cos\theta-1\right)
  d\phi
\end{equation}
%
So using (\ref{eq:comp}) to compare (\ref{A:nh}) with (\ref{sauron2b},\ref{sauron2}) we get
\begin{equation}
  \label{eq:p:vs:q}
  p={p^I} = {\frac{q}{2}}.
\end{equation}
Taking the same near horizon limit for the metric we obtain
\begin{equation}
  \label{eq:met:nh}
  ds^2= 2 d\tilde v d\tilde r+\frac{4L}{q}\tilde r d \tilde v d\vartheta'+L^2d\vartheta'^2
  +\frac{q^{2}}{4}\left[d\theta^{2}+\sin^{2}\theta d\phi^{2}\right]
\end{equation}
Let us for the moment consider the metric for constant $\theta$ and $\chi$. If we perform
the coordinate transformation
\begin{align}
  \label{eq:trans:1}
  d\vartheta'&=d\vartheta-\frac{q}{2 L}\frac{d\tilde r}{\tilde r}\\
  d\tilde v &= dt+\frac{q^2}{4}\frac{d\tilde r}{\tilde r^2}
\end{align}
we get
\begin{equation}
  \label{eq:II}
  ds^2= \frac{4L}{q}\tilde r dtd\vartheta+L^2d\vartheta^2+\frac{q^2}{4}\frac{d\tilde r}{\tilde r^2}
\end{equation}
Letting
\begin{equation}
  dt = dt'+\frac{q}{2}d\vartheta
\end{equation}
we obtain the more familiar form of BTZ
\begin{equation}
  \label{eq:btz}
  ds^2=\frac{4 L}{q}\tilde r dt d\vartheta +(L^2+2L\tilde r) d\vartheta^2
  +\frac{q^2}{4}\frac{d\tilde r}{\tilde r^2}
\end{equation}
Now defining
\begin{equation}
  \begin{array}{clcl}
    l&=q & \quad&\tilde r = \half(r^2-r_+^2)/(r_+) \\
    r_+ &= L&\quad &\tilde\phi=\vartheta+t'/l
  \end{array}
\end{equation}
we get the standard form of the BTZ metric
\begin{equation}
  ds^2=-\frac{(r^2-r_+^2)}{l^2r^2}dt'^2+\frac{l^2r^2}{(r^2-r_+^2)}dr^2
  +r^2\left(d\tilde\phi -\frac{r^2-r_+^2}{lr^2}dt'\right)^2
\end{equation}
Returning to (\ref{eq:II}) and letting
\begin{align}
  t&=l^2\tau/4, \\ \vartheta&=\psi/2 \\ e^0&= l/ L=q/L  \label{eq:def2}
\end{align}
we obtain
\begin{equation}
  \label{eq:final}
  ds^2=\frac{1}{4}{l}^{2}
  \left(
    -{\tilde r}^{2}{d\tau}^{2}+{\frac{{d\tilde r}^{2}}{{\tilde r}^{2}}}
  \right)
  +\frac{l^2}{4(e^0)^2}
  \left(
    d\psi+e^0\tilde rd\tau
  \right)^2
\end{equation}
This gives us the relationship between the $AdS_{2}$ and $S^{1}$ radii for the $AdS_{3}$
fibration.  To express this in terms of quantities in section~\ref{sec:br} we compare
(\ref{eq:final}) with our ansatz (\ref{sauron:ring2}), which gives:
\begin{align}
  w^{-1}v_{1}&=\tfrac{1}{4}l^{2},\\
  w^{2}&=\tfrac{1}{4}{l^{2}}{(e^{0})^{-2}}.
\end{align}
Upon eliminating $l^{2}$, one obtains the relation (\ref{eq:ads3:embedding}):
\begin{equation}
  w^{-1}v_{1}= w^{2}(e^{0})^{2},
\end{equation}
which is precisely what we obtained in section~\ref{sec:br} by solving the equation of
motion for $w$. 
This is analogous to the  Hopf fibration of $S^{3}$ whose metric can be written
\begin{equation}
  \label{eq:hopf:s3}
  ds^{2}=w^{-1}v_{2}(d\theta^{2}+\sin^{2}\theta d\phi^{2})+w^{2}(d\psi+p^{0}\cos\theta d\phi)^{2}
\end{equation}
with
\begin{equation}
   w^{-1}v_{2}= w^{2}(p^{0})^{2}.
\end{equation}

Finally, setting, $\tilde{p}^{0}=1$, and substituting (\ref{eq:ring:cijk},\ref{eq:p:vs:q},\ref{eq:def2}) into
(\ref{eq:E:special}) gives
\begin{equation}
  \label{eq:blah_final}
  {\cal E}= {4\pi^2\over  G_5}{(p^{3})}{(e^{0})^{-1}}=\frac{1}{4 G_{5}}(2\pi^{2}q^{2}L)=\frac{A_{H}}{4 G_{5}}
\end{equation}
which agrees with the result  in \cite{Elvang:2004ds}. 

\section{Spherically symmetric black hole near horizon geometry}
\label{sec:spherical:nh}

In this section, we find the near horizon geometry of a extremal spherically symmetric
black holes so that we can relate near horizon and asymptotic quantities. This will allow
us to compare (\ref{eq:E:special:2}) with known results.

We start with a spherically symmetric metric of the form
\begin{equation}\label{eq:ss:metric}
  ds^{2}=-f^{2}(\rho)d\tau^{2}+f^{-1}(\rho)(d\rho^{2}+\rho^{2}d\Omega_{(3)}^{2}).
\end{equation}
Assuming that we have an extremal black hole, near the horizon at $\rho=0$, $f$ will go like
\begin{equation}
  f(\rho)=\lambda \rho^{2} +{\cal O}(\rho^{3}).
\end{equation}
Now, expanding (\ref{eq:ss:metric}) to first non-trivial order in $\rho$, 
making the coordinate transformations
\begin{align}
  \label{eq:ss:coord1}
  \tau &= t/(2\lambda^{3/2}), \\
  \rho &= r^{1/2},\label{eq:ss:coord2}
\end{align}
and taking the near-horizon limit; $r\rightarrow\epsilon r$, $t\rightarrow t/\epsilon $,
$\epsilon\rightarrow0$; 
we can write the metric, (\ref{eq:ss:metric}), as
\begin{equation}
  \label{ss:nh}
  ds^{2}=\tfrac{1}{4}\lambda^{-1} \left[\left(-r^2dt^2 +{\frac{dr^2}{r^2}}\right)
    +\left(d\theta^2+\sin^2\theta d\phi^2\right) \right]
  +\tfrac{1}{4}\lambda^{-1}\left(d\psi+\cos\theta d\phi\right)^2.
\end{equation}
Comparing (\ref{ss:nh}) with (\ref{sauron1c}) (assuming $p^{0}=1$) we obtain
\begin{equation}\label{ss:w}
  w=\lambda^{-1/2}/2.
\end{equation}
Following the conventions of \cite{Kraus:2005zm}, the electric charge, $Q_{I}$, is given by
\begin{equation}
  \label{eq:ss:q}
  {H}_{IJ}F^{J\;\tau\rho}=  f\frac{Q_{I}}{\rho^{3}}.
\end{equation}
Now evaluating, (\ref{eq:ss:q}) near the horizon,  using
(\ref{eq:hvsf},\ref{eq:ss:coord1},\ref{eq:ss:coord2},\ref{ss:w}), gives
\begin{equation}
  \label{eq:ss:q2}
  \bar{f}_{IJ}F^{J}_{tr}=  \frac{1}{16}w^{-1}{Q_{I}}.
\end{equation}
Finally, recalling, $F^{J}_{tr}=e^{I}+a^{I}e^{0}$, and using (\ref{eq:q_Ihat}) we get
\begin{equation}
  16\hat{q}_{I}=Q_{I}
\end{equation}
as asserted in the text.

\sectiono{Non-supersymmetric ring near horizon geometry}
\label{sec:nonsusy}
\setcounter{equation}{0}
In section~\ref{sec:gen}, we construct the general entropy function for solutions with near horizon
geometries $AdS_2 \times U(1)^2$. Here, we begin with non-supersymmetric black ring
solution of \cite{Elvang:2004xi}, and show that it falls into the general class of
solutions mentioned in section~\ref{sec:gen}. Then we also evaluate the entropy of the black ring by
extremising the entropy function.  We consider the action
\begin{equation}
  I=\frac{1}{16\pi G_5}\int \sqrt{-g}\bigg( R-\frac{1}{4}F^2- \frac{1}{6\sqrt{3}}\;
  \epsilon^{\mu\alpha\beta\gamma\delta}A_\mu
  F_{\alpha\beta}F_{\gamma\delta} \bigg)\,,
\end{equation}
The metric for the non-SUSY solution is \cite{Elvang:2004xi}
\begin{align}
  \label{5dmetric}
  ds^2 &= -\frac{1}{h^2} \frac{H_{x}}{H_{y}} \frac{F_{y}}{F_{x}}
  \Big(dt+ A^0 \Big)^2\nonumber\\[3mm]
  & + h F_{x} H_{x} H_{y}^2\frac{R^2}{(x-y)^2}\Bigg[ - \frac{G_{y}}{F_{y} H_{y}^3}
  d\psi^2 - \frac{dy^2}{G_{y}} + \frac{dx^2}{G_{x}} + \frac{G_{x}}{F_{x} H_{x}^3} d\phi^2
  \Bigg]
\end{align}
The functions appearing above are defined as
\begin{align}
  F_{\xi} = 1 + \lambda \xi ,\qquad G_{\xi} = (1-\xi^2)(1+\nu\xi),\qquad H_{\xi} = 1- \mu
  \xi,
  \label{FGH}\end{align}
and
\begin{equation}
  \label{eq:h}
  h=1
  + {s^2 \over F_{x}H_{y} }(x-y)(\lambda+\mu)
\end{equation}
with
\begin{equation}
  s=\sinh\alpha\qquad c=\cosh\alpha
\end{equation}
The components of the gauge field are
\begin{align}
  A^1_t &=\sqrt{3}c /{h s} \;,
  \\
  A^1_\psi &=\sqrt{3}\frac{R(1+y)s}{h} \Bigg[ \frac{C_{\lambda}(c^2-h)}{s^2F_{y}} c^2
  -C_\mu\frac{3 c^2 -h }{H_{y}}
  \Bigg], \\
  A^1_\phi &=-\sqrt{3}\frac{R(1+x)c}{h} \Bigg[ \frac{C_{\lambda}}{F_{x}} s^2
  -C_\mu\frac{3c^2-2h}{H_{x}} \Bigg] ,
\end{align}
\begin{equation}
  \label{Cs}
  C_\lambda = \epsilon \sqrt{
    \lambda(\lambda+\nu) \frac{1+\lambda}{1-\lambda}},\qquad
  C_\mu = \epsilon  \sqrt{
    \mu(\mu+\nu)\frac{1-\mu}{1+\mu}}.
\end{equation}
A choice of sign $\epsilon = \pm 1$ has been included explicitly.  The components of the
one-form $A^0=A^0_\psi\; d\psi+A^0_\phi\; d\phi$ are
\begin{align}
  A^0_\psi(y) &= R(1+y)c \bigg[ \frac{C_\lambda}{F_{y}} c^2 - \frac{3C_\mu}{H_{y}} s^2
  \bigg]
  \\[2mm]
  A^0_\phi(x) &= - R\;\frac{1-x^2}{F_{x}H_{x}}\; \frac{\lambda+\mu}{1+\lambda}\; C_\lambda
  s^3\,,
\end{align}
The coordinates $x$ and $y$ take values in the ranges
\begin{equation}
  -1 \le x \le 1,\quad
  -\infty < y \le -1,\quad
  {{\mu^{-1}}} < y < \infty.
\end{equation}
The solution has three Killing vectors, $\partial_t$, $\partial_\psi$, and
$\partial_\phi$, and is characterised by four dimensionless parameters,
$\lambda, \mu ,\alpha$ and $\nu$, and the scale parameter $R$, which has dimension of
length.

Without loss of generality we can take $R>0$.  The parameters $\lambda, \mu$ are
restricted as
\begin{equation}
  0  \le \lambda < 1,\quad
  0 \le \mu < 1.
\end{equation}
The parameters are not all independent --- they are related by
\begin{equation}
  \label{eq:cond1}
  \frac{C_\lambda}{1+\lambda}s^2=\frac{3C_\mu}{1-
    \mu}c^2\,
\end{equation}
\begin{equation}
  \label{balance}
  \frac{1+\lambda}{1-\lambda}=\left( \frac{1+\nu}{1-\nu}\right)^2
  \left( \frac{1+\mu}{1-\mu}\right)^3.
\end{equation}
which, in the extremal limit, $\nu\rightarrow0$, implies
\begin{equation}
  \label{eq:l}
  \lambda=\frac{\mu(3+\mu^2)}{1+3\mu^2}
\end{equation}
and
\begin{equation}
  \label{eq:l2}
  s^2={3\over4}{(\mu^{-2}-1)}
\end{equation}
To avoid conical defects, the periodicities of $\psi$ and $\phi$ are
\begin{equation}
  \Delta\psi = \Delta\phi
  = 2 \pi {\sqrt{1-\lambda}}
  ({1+\mu})^{\frac{3}{2}}.
  \label{delphipsi}
\end{equation}

\subsection{Near horizon geometry}
\label{Nhg}

In the metric given by (\ref{5dmetric}),there is a coordinate singularity at $y=-1/\nu$
which is the location of the horizon. It can be removed by the coordinate
transformation~\cite{Elvang:2004xi}:
\begin{equation}
  \label{eq:coord:horizon}
  dt=dv+A^0_\psi(y)\,\frac{\sqrt{-F_{y}H_{y}^3}}{G_{y}} dy \, , \hspace{1cm} d\psi = d\psi' -
  \frac{\sqrt{-F_{y}H_{y}^3}}{G_{y}} dy.
\end{equation}
Letting, $\nu\rightarrow0$, making the coordinate change
\begin{equation}
  x=\cos\theta, \qquad y=-{R\over(\sqrt{\lambda\mu})\tilde r},
\end{equation}
and expanding to first non-trivial order in $\tilde r$, the metric becomes
\begin{align}
  ds^2&= [H_{x}K_{x}]\left( -{\tilde r^2 d\psi'^2} +2{\mu Rd\psi'd\tilde r }
    +{\mu^2R^2}{d\theta^2}
  \right)\nn\\
  &+\left[{\mu^2 K_{x}\over H_{x}^2F_{x}}\right]R^2{\sin^2\theta}d\phi^2
  +\left[{\lambda F_{x}H_{x}\over\mu K_{x}^2}\right] \left(dv+c_\psi (R
    d\psi')+A^0_\phi(x)d\phi\right)^2 +\ldots
\end{align}
where
\begin{equation}
  \label{eq:cpsi}
  c_\psi= c \left(
    \frac{C_{\lambda } c^2}{\lambda}
    +\frac{3  s^2 C_{\mu } }{\mu}
  \right)
\end{equation}
and
\begin{equation}
  \label{eq:lx}
  K_{x}= F_{x}+s^2(1+\lambda/\mu).
\end{equation}
We have neglected higher order terms in $\tilde r$ which will disappear when we take the
near horizon limit below.  Letting
\begin{align}
  \label{eq:psitilde}
  \tilde\psi&=\psi'+v/Rc_\psi\\
  u &= v/ R c_\psi\epsilon\\
  r&=\epsilon \tilde r
\end{align}
and taking, $\epsilon\rightarrow0$, the metric becomes
\begin{align}
  ds^2 &= [H_{x}K_{x}]\left( -{r^2 du^2} +{2\mu R du d r } +\mu^2R^2{d\theta^2}
  \right)\nonumber\\
  &+\left[\frac{\mu^2 K_{x}}{H^2_{x}F_{x}}\right]R^2{\sin^2\theta}d\phi^2+
  \left[\frac{\lambda F_{x}H_{x}}{\mu K^2_{x}}\right] \left(c_\psi (Rd \tilde \psi)+A^0_\phi
    d\phi\right)^2
\end{align}
Now we let
\begin{align}
  du &= du'+\frac{\mu R dr}{r^2}\\
  t &= \frac{u'}{\mu R}
\end{align}
Now we use the periodicities of $\phi$ and $\tilde \psi$ to redefine our coordinates,
\begin{equation}\label{rudra1}
  d \phi \rightarrow L d\phi,
  \qquad
  d \tilde \psi \rightarrow \frac{L d \tilde \psi}{2},
\end{equation}
where
\begin{equation}
  \label{eq:L:def}
  L=\sqrt{1-\lambda}(1+\mu)^{3/2}.
\end{equation}
Finally we can write the metric as in
(\ref{eq:anz:ef}),
\begin{align}
  \label{eq:ans:ef2}
  ds^2&=w^{-1}(\theta)\Omega^2(\theta)e^{2\Psi(\theta)}
  \left(-r^2dt^2+\frac{dr^2}{r^2}+\beta^2d\theta^2 \right)
  +w^{-1}(\theta)e^{-2\Psi(\theta)}(d\phi+e_\phi r dt)^2 \nonumber\\
  &+w^2(\theta)(d\psi+e_0 rdt+b_0(\theta)d\phi)^2\\
  A^1 &= e^1 r d t + b^1(\theta)(d\phi+e_\phi r dt) + a^1(\theta)(d\psi+e_0
  rdt+b_0(\theta)d\phi)
\end{align}
with,
\begin{align}\label{e13}
  \Omega &=\mu^{3/2}\lambda^{1/2} \frac{L^2}{2} c_{\psi}R^{3}\sin \theta,\\
  \label{e14}
  e^{-2 \psi}&=\frac{L^3 \mu^{3/2}\lambda^{1/2} c_{\psi}R^{3}\sin^2 \theta}{2 H_{x}^{3/2}F_{x}^{1/2}}\\
  \label{e141}
  e_{\phi}&= e_{0}=0  \\
  \label{e15}
  w &= \sqrt{\frac{L \lambda F_{x} H_{x}}{2 \mu}}\frac{c_{\psi}R}{K_{x}}\\
  \label{e16}
  b^0(\theta) &= \frac{2 A^0_{\phi}}{L c_{\psi}R}
\end{align}
The expression for the gauge fields reduce to,
\begin{align}\label{e161}
  A^1_t&=0\\
  \label{e162}
  A^1_{\psi}&=a^{1}(\theta)
  =\frac{\sqrt{3}R s}{h}\left(\frac{C_{\lambda}(c^2-h)c^2}{\lambda s^2}-C_{\mu}\frac{(3 c^2-h)}{\mu}\right)\\
  \label{e163}
  A^1_{\phi}&=b^1(\theta)+a^1(\theta)b^0(\theta)
  =-\frac{\sqrt{3}R c(1+\cos
    \theta)}{h}\left(\frac{C_{\lambda}s^2}{1+\lambda
      \cos \theta}-C_{\mu}\frac{(3 c^2-2h)}{1-\mu \cos \theta}\right)
\end{align}
and the expression for $b^1$ is,
\begin{align}\label{e163b}
  b^1(\theta)&=-\frac{\sqrt{3}R c(1+\cos \theta)}{h}\left(\frac{C_{\lambda}s^2}{1+\lambda \cos \theta}-C_{\mu}\frac{(3 c^2-2h)}{1-\mu \cos \theta}\right) \nonumber \\
  &-\frac{\sqrt{3}R s}{h}\left(\frac{C_{\lambda}(c^2-h)c^2}{s^2}-C_{\mu}(3
    c^2-h)\right)\frac{A^0_{\phi}(\cos \theta)}{c_{\psi}R}
\end{align}
where
\begin{equation}\label{e164}
  h = 1+s^2\frac{(\lambda + \mu)\cos \theta}{1+ \lambda \cos \theta}
\end{equation}
Then using the entropy function(4.5), the entropy of the non-supersymmetric black ring can
be expressed as,
\begin{equation}\label{192}
  \EE = 2 \pi^2 R^3[\mu^{3/2} \lambda^{1/2}(1-\lambda)(1+\mu)^{3}  (
    {C_{\lambda } c^2}/{\lambda}
    +{3  s^2 C_{\mu } }/{\mu})c ]
\end{equation}
which agrees with the extremal limit, $\nu\rightarrow0$, of Bekenstein-Hawking entropy of  the black-ring
solution in \cite{Elvang:2004xi}.

\bibliographystyle{JHEP} \bibliography{ring}

\end{document}